\documentclass[aip,pop,reprint,superscriptaddress,10pt]{revtex4-1}

\usepackage{graphicx}

\usepackage{amsmath}

\usepackage{color}

\usepackage{epstopdf}

\usepackage{multirow,bigstrut}

\usepackage{time,mathptmx,helvet}

\usepackage[
pdfstartview=XYZ,
bookmarks=false,
colorlinks=true,
linkcolor=blue,
urlcolor=blue,
citecolor=blue,
pdftex
]{hyperref}

\makeatletter

\newcommand{\Rmnum}[1]{\expandafter\@slowromancap\romannumeral #1@}
\makeatother

\begin{document}

\title{\large\textcolor{blue}{Ultrafast Proton Radiography of the Magnetic Fields Generated by a Laser-Driven Coil Current}\vspace{3pt}}

\author{Lan~Gao}
\altaffiliation{Current address: Princeton Plasma Physics Laboratory, Princeton University, Princeton, NJ, 08543, USA}
\affiliation{Department of Astrophysical Sciences, Princeton University, Princeton, NJ, 08544, USA} 

\author{Hantao~Ji}
\affiliation{Department of Astrophysical Sciences, Princeton University, Princeton, NJ, 08544, USA} 
\affiliation{Princeton Plasma Physics Laboratory, Princeton University, Princeton, NJ, 08543, USA} 

\author{Gennady~Fiksel}
\altaffiliation{Current address: Nuclear Engineering and Radiological Sciences (NERS), University of Michigan, Ann Arbor, MI, 48109, USA}
\affiliation{Laboratory for Laser Energetics, University of Rochester, Rochester, NY, 14623, USA} 

\author{William~Fox}
\affiliation{Princeton Plasma Physics Laboratory, Princeton University, Princeton, NJ, 08543, USA} 

\author{Michelle~Evans}
\affiliation{Laboratory for Laser Energetics, University of Rochester, Rochester, NY, 14623, USA} 
\affiliation{General Atomics, San Diego, CA, 92816, USA} 

\author{Noel~Alfonso}
\affiliation{General Atomics, San Diego, CA, 92816, USA}



\begin{abstract}

\noindent Magnetic fields generated by a current flowing through a U-shaped coil connecting two copper foils were measured using ultrafast proton radiography. Two $\sim$1.25 kJ, 1-ns laser pulses propagated through laser entrance holes in the front foil, and were focused to the back foil with an intensity of $\sim$3 $\times$ 10$^{16}$ W$/$cm$^{2}$. The intense laser-solid interaction induced a high voltage between the copper foils and generated a large current in the connecting coil. The proton data show $\sim$40-50 Tesla magnetic fields at the center of the coil $\sim$3-4 ns after laser irradiation. The experiments provide significant insight for future target designs that aim to develop a powerful source of external magnetic fields for various applications in high-energy-density science. 

\end{abstract}




\maketitle

\section{introduction}

While magnetic fields can be spontaneously generated in laser-driven high-energy-density (HED) systems, \cite{Stamper1971,Li2006,Willingale2010,Igumenshchev2014,Lancia2014,Gao2015} the capability of creating a strong external magnetic field source opens new opportunities in both basic and applied HED science.\cite{FikselRSI2015} Examples include creating magnetic reconnection geometries in laboratory, \cite{fiksel2014magnetic} magnetizing HED plasmas for fusion yield enhancement, \cite{Chang2011,PerkinsPoP} guiding hot electrons in fast ignition, \cite{Wang2015} and collimating positrons and electrons. \cite{HuiPOP2014} One technique that has attracted significant attention is based on laser irradiation of a metallic coil target. \cite{Daido1986,Courtois2005,fujioka2013kilotesla,Santos}

Such target is usually comprised of two parallel metallic foils, connected with a thin wire that is bent into different coil shapes for generating various magnetic field configurations. A high-energy laser passes through a hole in the front foil and irradiates the foil in the back. Hot electrons are generated during the intense laser-solid interaction \cite{Forslund} and escape from the back foil, \cite{PearlmanAPL} building up an electrical potential between the two foils.  This results in a large current flowing through the coil, and the generation of strong magnetic fields. 

The first attempt at applying this technique was carried out by Daido {\it et al} in 1986. \cite{Daido1986} Up to 60 T magnetic fields were measured at the center of a one-turn Cu coil irradiated with a CO$_{2}$ laser. Courtois  {\it et al}. used a Helmholtz configuration and measured a peak field of 7 T at the center of two Cu coils. \cite{Courtois2005} Recently, kilotesla magnetic fields were reported by Fujioka  {\it et al}. using a Ni coil target driven by a kJ-class laser. \cite{fujioka2013kilotesla} In these experiments, magnetic pickup probes or optical polarimetry were the primary diagnostics. Measurements of the field generation could only be made at $\sim$mm distances from the coil center. This is because magnetic probes are susceptible to electrostatic pickup contaminating the signal, and Faraday rotation can only measure field generation up to the critical density region associated with the optical probe beam wavelength. The magnetic field strength closer to the coil was inferred based on a vacuum model that describes the magnetic field profile along the probe path, which could lead to an overestimation of the total magnetic energy stored in the coil. \cite{fujioka2013kilotesla,Santos}

In this paper, direct measurements of the magnetic fields generated by a current flowing through a U-shaped Cu coil are reported using ultrafast laser-driven proton radiography. \cite{Borghesi_ppcf_ProtonImaging}  A high-energy proton beam was used to map the spatial distribution and time evolution of the magnetic fields around the coil. The experimental data show that the two parallel Cu foils provided a high voltage source induced by intense laser irradiation, driving a large current in the U-shaped coil. The measured current had an amplitude of $\sim$18-22 kA at $\sim$3-4 ns after laser irradiation, generating $\sim$200-250 T magnetic fields at the surface of the wire which spatially decayed to $\sim$40-50 T at the coil center. The conversion efficiency from laser energy to magnetic energy is $\sim$0.01-0.02$\%$ at these experimental conditions.

The paper is organized as follows. Section II describes the experimental setup. Section III presents a 3-D particle tracing model that guided the experiment. Section IV shows an analytical estimate of the proton deflections. Section V presents the experimental results and data interpretation. In Section VI the results are summarized.

\section{Experimental Setup}

\begin{figure}[t]
\includegraphics{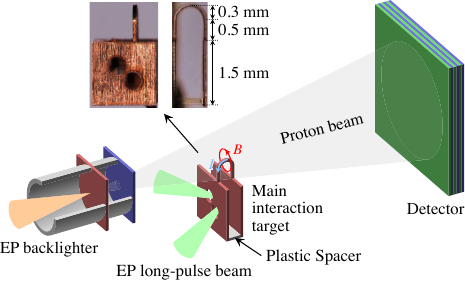}
\caption{\label{fig1} (color online). Experimental setup. Photographic views of an actual target from face-on and side-on directions are shown. The two Cu foils, 1.5 mm $\times$ 1.5 mm in size, are separated by 600 $\mu$m. The U-shaped coil has two 500-$\mu$m-long straight wires joined by a half-circular wire with curvature radius of 300 $\mu$m. A 125-$\mu$m-thick, 1.5 mm-long, and 600 $\mu$m-wide plastic spacer is sandwiched between the two Cu foils to maintain the foil planarity, whose location is clearly seen in the side-on view of the target. }
\vspace{-0.2cm}
\end{figure}

The experiments were carried out on the OMEGA EP Laser System at the University of Rochester's Laboratory for Laser Energetics. \cite{Waxer2005} Fig.~\ref{fig1} shows a diagram of the experimental setup. The main target was comprised of two parallel Cu foils (50-$\mu$m thick, 1.5 mm $\times$ 1.5 mm in area), connected by a U-shaped Cu coil with a wire cross section of 50 $\mu$m $\times$ 50 $\mu$m. The U-shaped coil had two 500-$\mu$m-long straight wires joined by a half-circular wire with a radius of curvature of 300 $\mu$m. The Cu foils were separated by 600 $\mu$m. The front foil had two laser entrance holes, 200 $\mu$m in diameter. The target was laser cut from a Cu foil, and bent around a circular rod to the required shape. A 125-$\mu$m-thick plastic spacer (1.5 mm $\times$ 600 $\mu$m in area) was sandwiched between the Cu foils at their bottom to maintain the foil position. Face-on and side-on photographic views of an actual target are shown in Fig.~\ref{fig1}. The target dimensions were accurately measured before the target was fielded in the experiment.

Two OMEGA EP long-pulse beams, each propagating through a laser entrance hole, were focused to a $\sim$100-$\mu$m-diam focal spot at 23$^{\circ}$ angle of incidence to the back foil. Each long-pulse beam delivered $\sim$1.25-kJ of energy in a 1-ns square temporal profile at a wavelength of 351 nm, corresponding to an overlapped laser intensity of $\sim$3 $\times$ 10$^{16}$ W$/$cm$^{2}$. 

As the long-pulse lasers irradiate the back Cu foil, a beam of hot electrons with a temperature of tens of keV is generated. \cite{Forslund}  These electrons have enough energy to escape from the target, building up a positive potential on the back foil. \cite{Sinenian2013} A fraction of these hot electrons are captured by the front foil within tens of picoseconds. These processes result in the generation of a large current in the U-shaped coil that connects the Cu foils. The current keeps growing during the laser irradiation, corresponding to the charging of the target and the subsequent discharging in the U-shaped coil, with a rise time comparable to the laser pulse duration. \cite{Daido1986,Santos} After the laser pulse terminates, the target behaves like an inductance-resistance electrical circuit, enabling the current to be sustained for tens of nanoseconds. \cite{Santos}

In these experiments, the laser-driven current in the coil was measured with an ultrafast proton beam in a face-on geometry (see Fig.~\ref{fig1}). Up to 55 MeV protons were generated by irradiating a 20-$\mu$m-thick Cu foil with the 0.3-kJ, 1-ps OMEGA EP backlighter beam at a wavelength of 1.053-$\mu$m due to the target normal sheath acceleration mechanism. \cite{wilks2001} The intense laser was focused with a 1-m focal length, f/2 off-axis parabolic mirror onto the Cu target at normal incidence, providing an intensity of 1.5 $\times$ 10$^{19}$ W$/$cm$^{2}$. A thin Ta foil was used to protect the Cu foil from coronal plasma and x-ray photons created in the main interaction. \cite{zylstra:013511, Gao2013} The proton beam provides a spatial resolution of 5 to 10 $\mu$m and a temporal resolution of a few picosecond.

\begin{figure}[t]
\includegraphics{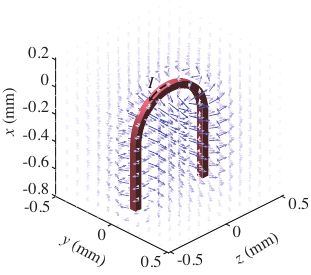}
\caption{\label{fig2} (color online). Simulated magnetic field distributions by a U-shaped coil current, with coil geometry taken from the actual target. The protons are incident in $z$ direction. }
\vspace{-0.2cm}
\end{figure}

The protons were detected with a filtered stack of radiochromic film, providing two-dimensional images of the interaction. \cite{Borghesi_ppcf_ProtonImaging} The system magnification $M$, defined as $M=(D + d)/d$, where $d$ was the distance from the proton-source foil to the middle of the capacitor and $D$ was the distance from the middle of the capacitor to the radiochromic film, was $\sim$ 12.5 to 15. Each film corresponds to a proton energy determined by proton energy deposition inside the detector at which the Bragg peak occurs, thereby diagnosing the long-pulse interaction at times based on the proton time-of-flight to the main target and the timing difference between the long- and short-pulse beams. The relative timing between the long-pulse and the short-pulse beams was measured with an x-ray streak camera.

\section{Particle ray-tracing model}

To provide insight for experimental design and data interpretation, a 3-D particle ray-tracing program was developed to simulate the proton trajectories in prescribed electromagnetic fields.

The proton beam was generated at the location of the backlighter Cu foil and propagated through the coil target. The amplitude and distribution of the 3-dimensional magnetic fields generated by the current in the U-shaped coil were calculated using the Biot-Savart law. No electrical fields were included. As energetic protons probed through the field region, the proton beam spatial profile underwent variations due to deflections from the Lorentz force. Synthetic proton images were constructed by tracing each proton trajectory and accumulating proton numbers at the detector plane. The radiography geometry was taken from the experiment. The current and proton energy were free parameters in the simulations.

Figure~\ref{fig2} shows the magnetic field distribution generated by a U-shaped coil current. The coil geometry is the same as the target that was used in the experiment. The blue arrow represents the magnetic field vector, with its length representing the field strength. Strong poloidal magnetic fields are shown to wrap around the half-circular wire and the two straight parts of the coil.

\begin{figure}[t]
\includegraphics{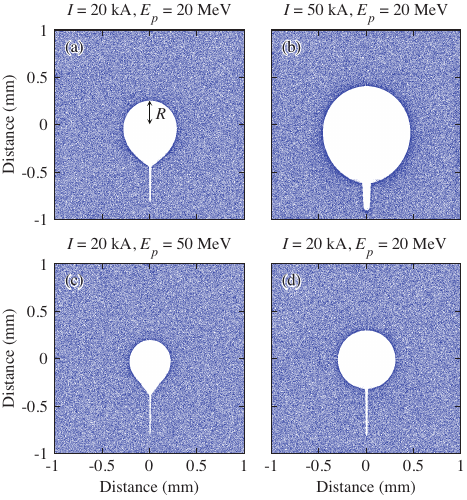}
\caption{\label{fig3} (color online). Predicted proton radiographs calculated from the particle tracing simulations for various currents and proton energies. (a), (b) and (c) are for a U-shaped coil current. (d) is calculated for the case where the connecting wire is straight.}
\end{figure}

Synthetic proton images calculated for various currents and proton energies are shown in Fig.~\ref{fig3}. $I$ represents the current amplitude and $E_{p}$ is the proton energy. Each image was calibrated from the detector plane to the object plane taken the system magnification into account. A proton void is seen in the center of the simulated radiograph for all the U-shaped coils, indicating complete evacuation of protons in this region. Each void has a sharp boundary indicating formation of caustics. \cite{KuglandRSI,levy2015development} These features result from the path-integrated proton deflection by the magnetic fields generated from the current in the half-circular wire. Because magnetic fields generated by straight sections of the wires on either side of the U are approximately parallel to the proton trajectories, little proton deflection is observed at these locations. Proton stopping inside the Cu wires is therefore dominant. As a test, a straight wire connecting the two separated wires is also simulated. Symmetric azimuthal magnetic fields are generated around the wire, resulting in a circular void in the proton image, as is shown in Fig.~\ref{fig3}(d). 

\begin{figure}[t]
\includegraphics{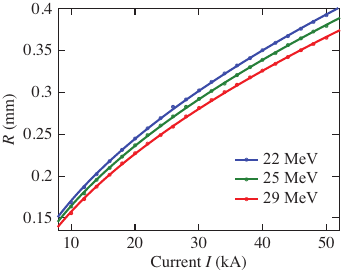}
\caption{\label{fig4} (color online). Simulated {\it{R}} as a function of current $I$. The blue, green, and red line represents 22 MeV, 25 MeV and 29 MeV protons respectively.}
\end{figure}

The size of the void increases for larger currents [Fig.~\ref{fig3}(a) and (b)], and decreases for higher proton energies [Fig.~\ref{fig3}(a) and (c)]. Because of the irregularity of the prolate void structure, the distance between the U-shaped coil apex at the coordinate system origin to the prolate void apex, designated as {\it{R}} in Fig.~\ref{fig3}(a), was used as one way of characterizing these effects. Fig.~\ref{fig4} shows the simulated {\it{R}} as a function of coil current for 29 MeV, 25 MeV and 22 MeV protons. Power fittings of these lines show that {\it{R}} is proportional to $I^{0.5}$, and inversely proportional to $E_{p}^{0.25}$.

There are other methods to characterize proton deflections, such as the average size of the prolate void or the distance between the coil apex to the lower side of the void structure. In experimental data, presence of dark filaments made it difficult to identify the lower foil-side boundary (see Section V). The void size in the upper side {\it{R}}, was therefore a robust parameter to be used for inferring current amplitude and its induced magnetic fields.

\section{Analytical Estimate}

Caustic structure in the proton image can be mathematically identified by basic paraxial mapping, between a proton ray at the object plane and the same ray at the detector plane after deflection by electromagnetic fields. \cite{KuglandRSI} In this section, a simplified model is derived to show analytically the most important effects near the coil, namely formation of a proton void with a sharp caustic boundary. 

The proton rays, generated at the location of the backlighter foil as a point source, pass through the magnetic field region which is located at a distance $d$ from the proton source. Each proton ray is deflected by the field with an angle of $\alpha$. The protons then propagate ballistically until reaching the detector at a distance $D$ from the field region. The coordinate used here is ($x$, $y$, $z$) with the origin situated at the coil center, and the $z$-axis passing through the proton source and the detector. $z = -d$ is therefore the proton generation plane or the object plane, and $z = D$ is the detector plane or the image plane. 

A particular proton ray is characterized by ($x$, $y$), the interaction point of the proton ray with the object plane. After deflected with an angle of $\alpha (x,y)$ by the fields, the proton ray reaches the image plane at the point ($x_{i}$, $y_{i}$). The point ($x$, $y$) is mapped to the point ($x_{i}$, $y_{i}$) following
\begin{eqnarray}
x_{i} &=& x + D\frac{x}{d} + D\alpha_{x} = Mx + D\alpha_{x} ,  \\ [0.25cm]  
y_{i} &=& y + D\frac{y}{d} + D \alpha_{y} = My + D\alpha_{y}.   
\end{eqnarray}

The deflection angle of protons from their original trajectories while passing through magnetic field $\mathbf{B}$ can be calculated within the paraxial approximation as
\begin{eqnarray}
\alpha_{x} &\approx& \frac{v_{x}}{v_{z}} = - \frac{e}{mv_{z}}\int{B_{y}}dz = - \frac{e}{\sqrt{2mE_{p}}}\int{B_{y}}dz,\\[0.25cm]
\alpha_{y} &\approx& \frac{v_{y}}{v_{z}} = \frac{e}{mv_{z}}\int{B_{x}}dz = \frac{e}{\sqrt{2mE_{p}}}\int{B_{x}}dz.
\end{eqnarray}
Here $e$ is the proton charge, $m$ is the proton mass, and $B_{x}$ and $B_{y}$ are the $x$ and $y$ component of the $\mathbf{B}$ field. 

We now simplify our system to calculate the proton trajectories passing through a straight wire parallel to the $z$-axis with a length of $L_{z}$.  This straight segment is qualitatively similar to the half-circular wire of the U in the experiment.

As a first-order estimate, the $\mathbf{B}$ field from this straight wire is assumed to have a distribution function same as that of an infinitely long current:
\begin{equation}
|B(r)| = \frac{\mu_{0}I}{2\pi r} \quad (r\ge R_{0}).
\end{equation}
Here $R_{0}$ is the radius of the coil, and $r$ is the distance from the proton ray to the current, which equals to $\sqrt{x^2+y^2}$. This leads to
\begin{eqnarray}
B_{x}(r) =&  B(r)\displaystyle\frac{y}{r} = \displaystyle\frac{\mu_{0}I}{2\pi}\frac{y}{r^{2}}       \quad    (r\ge R_{0}),\\ [0.25cm]
B_{y}(r) = &-B(r)\displaystyle\frac{x}{r} = - \displaystyle\frac{\mu_{0}I}{2\pi}\frac{x}{r^{2}}      \quad   (r\ge R_{0}).
\end{eqnarray}

 This system now has a cylindrical symmetry, where the proton deflections depend only on $r$. The mapping relation between the object plane  ($r$, $z$) and the image plane ($r_{i}$, $z_{i}$) can be found as
\begin{eqnarray}
r_{i} = Mr+\frac{\mu_{B}\Delta zD}{r}  \quad  (r\ge R_{0}),
\end{eqnarray}
where $\mu_{B} = \displaystyle\frac{e\mu_{0}I}{2\pi\sqrt{2mE_{p}}}$, and $\Delta z = \displaystyle\int{dz}$ which is the proton path length integrated over the field region. Eq. (8) shows that $r_{i} >  MR_{0}$, and therefore a proton void is formed in the proton image.

When $dr_{i}/dr=0$, a caustic is formed, creating a spatially sharp pile-up of protons in the image intensity. This caustic corresponds to the edge of the proton void and determines its size. Solving the equation $dr_{i}/dr=0$ leads to 
\begin{eqnarray}
r^* = \sqrt{\frac{\mu_{B}\Delta zD}{M}}.
\end{eqnarray}
The corresponding proton void radius on the image plane is 
\begin{eqnarray}
r_{i} = 2 \sqrt{\mu_{B}\Delta zDM}.
\end{eqnarray}
Taken the system magnification account, the void radius at the object plane is 
\begin{eqnarray}
r_{M} = 2\sqrt{\frac{\mu_{B}\Delta zD}{M}}.
\end{eqnarray}
which is proportional to $I^{0.5}$, and inversely proportional to $E_{p}^{0.25}$. This is in good agreement with the scaling law calculated by the ray-tracing simulations.

For 20-MeV protons and a 20-kA current, the proton void radius estimated from Eq. (11) is 0.31 mm assuming the proton path length equal to the length of the connecting coil $L_{z}$. This is also consistent with that calculated in simulations, as is shown in Fig.~\ref{fig3}(d).

\section{Experimental Results}

\begin{figure*}[t]
\includegraphics{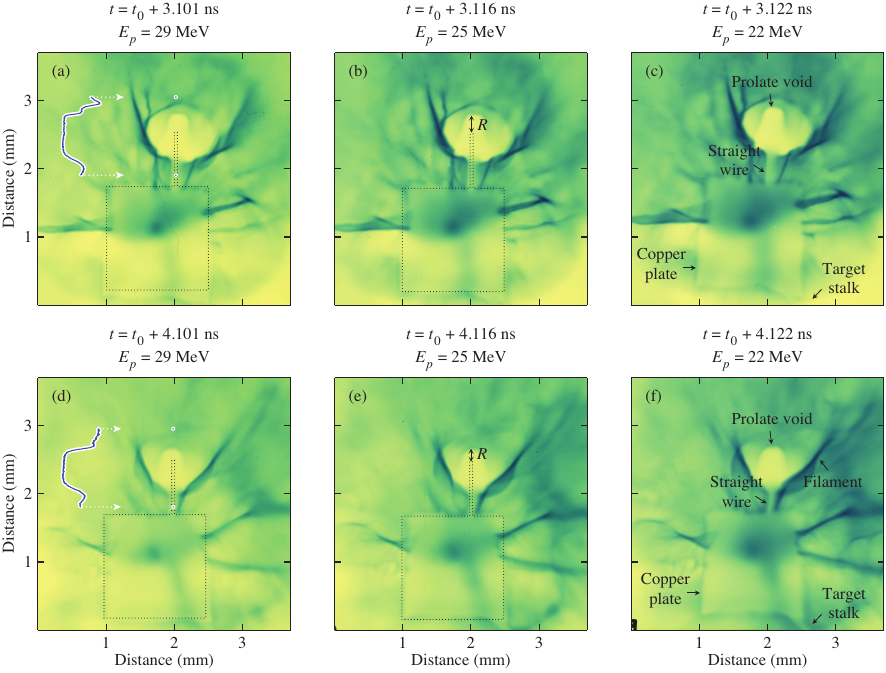}
\caption{\label{fig5} (color online). Proton radiographs obtained with 29-MeV, 25-MeV and 22-MeV protons at six different times. The data in two rows are obtained from two different shots. Overlaid on top of (a), (b), (d) and (e) is the contour of the original target in a face-on view. Lineouts drawn along the U-shaped coil apex are shown in (a) and (d).}
\end{figure*}

Numerical simulations and analytical calculations have shown that a void structure with full evacuation of protons is formed in the proton image for the current-generated magnetic fields. This void is around the coil apex and has a sharp boundary. These features have been used for the interpretation of experimental results.

Figure~\ref{fig5} shows six raw proton radiographs of the target after laser irradiation. (a), (b) and (c) are obtained from the first shot, where the short-pulse beam for generating energetic protons was fired 1 ns earlier than in the second shot, from which (d), (e) and (f) are collected. The exact proton probing time is shown for each image with respect to $t_{0}$, where $t_{0}$ is the arrival time of the long-pulse beams at the surface of the back Cu foil. The proton energy was 29 MeV, 25 MeV and 22 MeV for the three columns respectively. The radiographic view of the target corresponded to 3.7 mm $\times$ 3.7 mm at the object plane. The data show the rectangular Cu foil, the fiber stalk that holds the entire target, and the straight-part of the U-shaped coil. Overlaid on top by the dashed line on images (a), (b), (d) and (e) is the contour of the original target in a face-on view, in good agreement with the experimental measurement.

Around the U-shaped coil apex, a proton void without proton accumulations is observed for all six images, revealing magnetic field generation induced by the current in the U-shaped coil. At the void edge, a sharp increase in proton numbers is also seen, indicated by the lineouts in Fig.~\ref{fig5}(a) and (d).  This void structure is not caused by electric fields or wire expansion due to resistive heating, as no such proton deflections are observed for the straight portions of the wire joining the U to the foils.

Dark filamentary structures are also observed in the proton images, indicating a high proton flux accumulation along each striation. These structures, possibly associated with instabilities in the coronal plasma, \cite{Haines1981} originate from the ablation region of the target and extend into the low-density region under the confining force of the self-generated electromagnetic fields. \cite{Filaments, Gao2012,WILLI1981,Nilson_PoP} At these probing times, the coronal plasma produced by the main interaction between the long-pulse beams and the back Cu foil had arrived at the coil region, contributing to the formation of filaments in the proton images. The filamentary structures are more prevalent in the upper half of the image, as plasma expansion into the lower part is stopped by the plastic spacer between the Cu foils (location of the spacer can be found in Fig.~\ref{fig1}). 

Furthermore, Figs.~\ref{fig5}(a)-(c) show a circular structure around the wire. This feature is also excluded as due to the coil current-induced magnetic fields, because no complete void is exhibited. 

\begin{figure}[t]
\includegraphics[width=8.6cm]{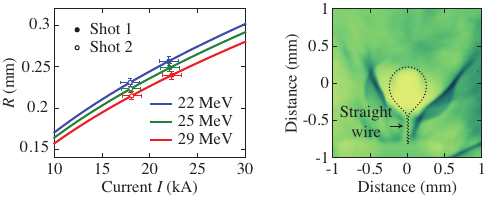}
\caption{\label{fig6} (color online). Left panel: simulated and measured {\it{R}} as a function of current. The blue, green, and red line represents 22 MeV, 25 MeV and 29 MeV proton energy respectively. The dots were calculated from the first shot data, and the circles were from the second shot. Right panel: synthetic proton radiograph calculated from the particle tracing simulations with 18-kA current in the coil is overlaid on top of the measured data reproduced from Fig.~\ref{fig5}(e).}
\end{figure}

The amplitude of the current in the U-shaped coil and the current-generated magnetic fields are estimated by matching the theoretically calculated and measured proton voids in the proton images. To determine the evolution of the void size {\it{R}}, lineouts are drawn for all six images along the direction passing through the U-shaped coil apex (see Fig.~\ref{fig5}(a) and (d) for examples). The left panel in Fig.~\ref{fig6} shows the measured values of {\it{R}} overlaid on top of the simulated results. The three dots correspond to the measurements from Figs.~\ref{fig5}(a)-(c), where the data were taken from the first shot. The three circles are measured from Figs.~\ref{fig5}(d)-(f) taken from the second shot. Within the same shot, the relative proton probing time spans about 21 ps, as the proton energy is changed from 22 MeV to 29 MeV. The inferred current amplitude from each measured {\it{R}} remains about the same, indicating a static current during the 21 ps interval. In time, the amplitude of the current decreased, from 22 kA at about $t=t_{0}+3.1$ ns to 18 kA at about $t=t_{0}+4.1$ ns. The error bar in current is $\pm$1 kA, as a result of uncertainty of $\pm$5 $\mu$m in determining {\it{R}}.

The inferred current in the coil is supported by comparing the simulated void structure with the experimental measurement. The right panel in Fig.~\ref{fig6} shows the simulated proton structure for 25 MeV protons with the current amplitude of 18 kA, overlaid on top of experimental data reproduced from Fig.~\ref{fig5}(e). The experimental data also shows two dark filaments originating from either side of the straight coil and deviating from their original paths while extending further into the coil region, making it difficult to exactly identify the lower foil-side boundary of the void and compare with simulations in which plasma effects are not included. The upper side where less plasma effects exist, however, is better reproduced by simulations with 18 kA current. 

The corresponding magnetic field generated by 18-22 kA current at the surface of the wire is $\sim$200-250 T, spatially decaying to $\sim$40-50 T at the center of the half-circular wire. The magnetic energy stored in the U-shaped coil is $\sim$0.26-0.4 J, leading to a conversion efficiency of $\sim$0.01-0.015$\%$ from laser energy to magnetic energy. 

Electrical fields or wire expansion due to resistive heating might play a role in deflecting protons propagating through the straight-part of the coil. Proton deflections by magnetic fields around the straight wires are small, because the incident protons are close to parallel to the magnetic field lines in this region. Assuming that the measured displacement of $\sim$25 $\mu$m for 25-MeV protons is caused by electrical fields, an upper-bound estimation of the electrical field strength is $\sim$5 $\times$ $10^{8}$ V$/$m. 

\begin{figure}[t]
\includegraphics{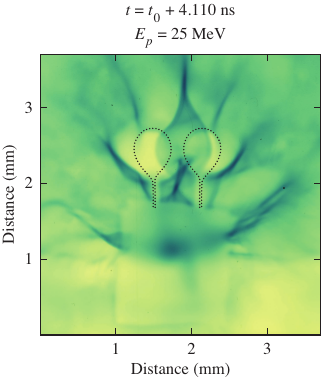}
\caption{\label{fig7} (color online). Proton radiograph obtained with 25-MeV protons at $t=t_{0}+4.110$ ns for a target with two U-shaped coils. The two coils are separated by 600 $\mu$m. The calculated proton image for this case is overlaid on top, shown in dashed lines.}
\end{figure}

Further experiments with two connecting U-shaped coils forming a Helmholtz-like configuration were carried out, demonstrating the two Cu foils provided a voltage source rather than a current source for driving the current in the U-shaped coils. Fig.~\ref{fig7} shows a proton radiograph for these targets driven by the same laser conditions as the single coil case. The data was obtained with 25-MeV protons at $t=t_{0}+4.110$ ns. The two coils were separated by 600 $\mu$m. Compared to the case with a single coil (Fig.~\ref{fig5}(e)), two void structures are observed, each created by a U-shaped coil current. In order to reproduce the observed structures, 18 kA of currents in both coils have to be prescribed in the ray tracing program, identical to the current flowing in the single-coil experiment. This indicates that under these conditions the target acts as a voltage source with output impedance much smaller than the coil inductive or resistive impedance. The conversion efficiency from laser energy to magnetic energy is increased to $\sim$0.02$\%$, providing a possibility for increasing the stored magnetic energy by adding extra coils. It is noted that a jet-like feature is observed in between the two ring structures, whose formation is under investigation and will be described in a future publication. 

\section{Conclusion}

In conclusion, direct measurements of the magnetic fields generated by a current flowing through a Cu coil target driven by high-energy lasers at a focused laser intensity of $\sim$3 $\times$ 10$^{16}$ W$/$cm$^{2}$ are reported using ultrafast proton radiography. The measured current amplitude was $\sim$22 kA at about 3.1 ns after laser irradiation, which decayed to $\sim$18 kA in 1 ns. The corresponding magnetic fields induced by the current were $\sim$200-250 T at the surface of the wire, spatially decayed to $\sim$40-50 T at the coil center. The conversion efficiency from laser energy to magnetic energy was $\sim$0.01-0.02$\%$ at these experimental conditions. 

Besides observing the proton void due to the current-induced magnetic fields, dark filamentary structures were seen in all the images. This is because the coronal plasma expanded into the coil region and the self-generated electromagnetic fields associated with instabilities inside the plasma caused proton deflections. Creation of such features overlap with the proton void, complicating the experimental data interpretation. To avoid this issue, future target design could have longer straight-part of the coil so that it takes longer time for the coronal plasma to reach the region of interest, or include a shielding piece to prevent the coronal plasma from expanding into the region.

The experimental data show that the Cu foils provided a voltage source for driving the current in the connecting coil. This indicates that the load resistance and inductance will play a role in the ultimate current. Detailed experiments will be carried out to develop lumped circuit models for these laser-driven systems and understand the underlying physics for controlling the field dynamics. This will benefit the proposed applications that use this technique for external magnetic fields.

\section*{Acknowledgments}

This work was supported by the National Laser Users Facility under Grant No. DE-NA0002205. The authors express their gratitude to J. Y. Zhong and Y. T. Li for providing ideas on target design, to Q. L. Dong, P. Nilson, and K. Hill for useful discussions, to General Atomics and the Laboratory for Laser Energetics (LLE) for target fabrication, and to the OMEGA EP crew for technical support. 


\bibliography{1MAIN}

\end{document}